*Research Article*

# Canonical PSO Based $K$-Means Clustering Approach for Real Datasets

**Lopamudra Dey[1] and Sanjay Chakraborty[2]**

[1] *Heritage Institute of Technology, Kolkata, West Bengal 700 107, India*
[2] *Institute of Engineering & Management, Kolkata, West Bengal 700 091, India*

Correspondence should be addressed to Sanjay Chakraborty; sanjay_ciem@yahoo.com

"Clustering" the significance and application of this technique is spread over various fields. Clustering is an unsupervised process in data mining, that is why the proper evaluation of the results and measuring the compactness and separability of the clusters are important issues. The procedure of evaluating the results of a clustering algorithm is known as cluster validity measure. Different types of indexes are used to solve different types of problems and indices selection depends on the kind of available data. This paper first proposes Canonical PSO based $K$-means clustering algorithm and also analyses some important clustering indices (intercluster, intracluster) and then evaluates the effects of those indices on real-time air pollution database, wholesale customer, wine, and vehicle datasets using typical $K$-means, Canonical PSO based $K$-means, simple PSO based $K$-means, DBSCAN, and Hierarchical clustering algorithms. This paper also describes the nature of the clusters and finally compares the performances of these clustering algorithms according to the validity assessment. It also defines which algorithm will be more desirable among all these algorithms to make proper compact clusters on this particular real life datasets. It actually deals with the behaviour of these clustering algorithms with respect to validation indexes and represents their results of evaluation in terms of mathematical and graphical forms.

## 1. Introduction

One of the best known problems in the data mining is the clustering. Clustering is the task of categorising objects having several attributes into different classes such that the objects belonging to the same class are similar, and those that are broken down into different classes are not [1]. There are several clustering algorithms that have been proposed till now. Due to no prior information in clustering, the suitable evaluation of the results is necessary. Evaluation means measuring the similarity between clusters, measuring the compactness, and separation between clusters [2]. Evaluation measurement is also proposed as a key feature in internal and external cluster validation indexes [3]. Such a measure can be used to compare the performance of different data clustering algorithms on different real life datasets. These measures are usually tied to the type of criterion being considered in assessing the quality of a clustering method. Three different techniques are available to evaluate the clustering results: external, internal, and relative [4]. Both internal and external criteria are based on statistical methods and they have high computation demand. The external validity methods evaluate the clustering based on some user specific intuitions [4]. The objective of this paper is the comparison of the different clustering schemas that have been already proposed [5] with Canonical PSO based $K$-means clustering algorithm.

The rest of the paper is organized as follows. The Canonical PSO based $K$-means algorithm is proposed in Section 2 with some other existing clustering algorithms. Some popular and widely used validity indices are introduced in Section 3. Section 4 demonstrates the clustering compactness measurements on a toy example dataset using $K$-means and DBSCAN clustering algorithms. Section 5 demonstrates the clustering compactness measurements with experimental results and comparison of the indices is outlined in this section, and Section 7 gives a brief conclusion of this paper. Interested readers may found some significant references at the end section of this paper.

This paper mainly focuses on the implementation of Canonical PSO based *K*-means algorithm. Some cluster validity indices (e.g., Dunn index, DB index, Silhouette index, Rand index, Mirkin index, etc.) and ANOVA test are analysed. All these indices are individually experimented on the air pollution dataset, customer, wine, and vehicle dataset using typical *K*-means, DBSCAN, Hierarchical, simple PSO based *K*-means, and Canonical PSO based *K*-means algorithms. The overall motivation of this paper is given in Figure 2.

## 2. Clustering Algorithms

This section introduces a new modified Canonical PSO based *K*-means clustering algorithm and also describes all these well-known algorithms briefly.

*2.1. Simple PSO Based K-Means Clustering.* James Kennedy and Russell C. Eberhert originally proposed the Particle Swarm Optimization (PSO) algorithm for optimization. It is a population based, robust, and stochastic optimization technique mainly designed for the balancing weights in neural networks [6]. In data clustering, it is possible to view the clustering problem as an optimization problem that locates the optimal centroids of the clusters rather than to find an optimal partition. This view offers us a probability to apply PSO optimal algorithm on the clustering solution. The use of basic PSO technique in document clustering analysis is proposed in many papers [7, 8].

In the PSO based *K*-means algorithm, the capability of globalized searching of the PSO algorithm and the fast convergence of the *K*-means algorithm are combined [9]. The algorithm results in better accuracy than existing algorithms. In the original PSO, at any instance each particle has a position and a velocity. At the beginning, population of particles is initialized with random positions denoted by vectors $x_i$ and random velocities $v_i$. The population of such particles is called a swarm, *S*. Each particle is searching for the optimum. Each particle remembers the position it was in where it had its best result so far (its personal best) [7]. The particles in the swarm cooperate. They exchange information about what they have discovered in the places they have visited.

In each time step, a particle has to move to a new position. It does this by adjusting its velocity. Velocity is updated based on information obtained in previous steps of the algorithm.

This updating of velocity and position can be described by the following set of equations:

$$v_{ij}(t+1) = v_{ij}(t) + C1R1\left(p_{ij}(t) - x_{ij}(t)\right) \\ + C2R2\left(p_{ij}(t) - x_{ij}(t)\right), \quad (1)$$

$$x_{ij}(t+1) = x_{ij}(t) + v_{ij}(t), \quad (2)$$

where $i = 1, 2 \ldots N$, $j = 1, 2 \ldots n$.

Here, $v_{ij}(t+1)$ is the new velocity at time step $(t+1)$, $v_{ij}(t)$ is the old velocity at time step $t$, $p_{ij}(t)$ is the best position of each particle, $p_{gj}(t)$ is the best position of swarm, $x_{ij}(t+1)$ is current position of each particle, and $x_{ij}(t)$ is old position of each particle.

*R*1 and *R*2 are random variables uniformly distributed within [0, 1] and *C*1*C*2 are weighting factors, also called the cognitive and social parameter, respectively. In the first version of PSO, a single weight, $C = C1 = C2$, called acceleration constant, was used instead of the two distinct weights in (1).

*2.2. Canonical PSO Based K-Means Clustering.* Oscan and Mohan (1999) focused on the early PSO model of (1) and (2), and they showed that particles were actually moving on sinusoidal waves per coordinate of the search space, while causticity was offering a means to control its frequency and amplitude. They had modified the PSO model and the model is defined by the following equation:

$$v_{ij}(t+1) = \chi \left[ v_{ij}(t) \\ + C1R1\left(p_{ij}(t) - x_{ij}(t)\right) \\ + C2R2\left(p_{gj}(t) - x_{ij}(t)\right)\right], \quad (3)$$

$$x_{ij}(t+1) = x_{ij}(t) + v_{ij}(t),$$

where $i = 1, 2, \ldots, N$ particles (vector data) and $j = 1, 2, \ldots, n$.

Where chi($\chi$) is a parameter called constriction coefficient or constriction factor, responsible for keeping the particle moving in the same direction, it was originally heading, while the rest of the parameters remain the same as for the previously described PSO models [9].

*2.3. Proposed Algorithm*

*Input*

    *K* = number of clusters.

    *N* = number of iterations.

    Data = different real dataset.

*Output.* A set of clusters.

(1) Initialize each particle randomly by taking *k* different data samples from the dataset as the initial cluster centers.

(2) Initialize velocity and personal best position of each particle.

(3) Repeat for each particle.

    (a) Calculate the performance of each particle based on the *k*-means fitness function

$$J(K) = \sum_{i=1}^{k} D^2\left(v_i, x_j\right), \quad (4)$$

where $x_j$ denotes the $j$th data point, $v_i$ denotes the center of the $i$th cluster $C_i$, and $D(v_i, x_j)$ denotes the distance (e.g., Euclidean distance) of $x_j$ from $v_j$.

(b) Reassign the data vector to the centroid vector according to the fitness value.

(c) Update the personal best position.

(d) Update the global best position.

(e) Modify velocity and position of each particle using formula (3) and generate the next solution.

(4) Repeat Step (3) until one of the following termination conditions is satisfied.

(a) The maximum number of iterations is exceeded.

(b) The average change of centroid vector is less than the predefined value.

*2.4. K-Means Algorithm.* $K$-means clustering is a method of cluster analysis which aims to partition $n$ observations into $K$ clusters depending on some similarity/dissimilarity metric where the value of $K$ may or may not be known a priori [7]. The objective of $K$-means clustering algorithm is usually to create one set of clusters that partitions the data into similar groups. As a result, maximum similarity samples are placed in same cluster and low similarity samples are placed in different clusters [10].

*2.5. Hierarchical Algorithm.* In hierarchical clustering clusters are generated by grouping data with similar pattern of expression across a range of samples located near each other. Hierarchical clustering calculates all pairs-wise distance associations between samples and experiments to merge pairs of values that are mainly similar [9].

*2.6. DBSCAN Algorithm.* This is a density-based clustering algorithm that produces a partitional clustering, in which the number of clusters is automatically determined by the algorithm. DBSCAN requires two parameters: $\varepsilon$ (eps) and the minimum number of points required to form a dense region (minpts) [11]. It starts with an arbitrary starting point that has not been visited. This point's $\varepsilon$-neighbourhood is retrieved, and if it contains sufficiently many points, a cluster is started. Otherwise, the point is labelled as noise. If a point is found to be a dense part of a cluster, its $\varepsilon$-neighbourhood is also part of that cluster [12]. Hence, all points that are found within the $\varepsilon$-neighbourhood are added as is their own $\varepsilon$-neighbourhood when they are also dense. This process continues until the density-connected cluster is completely found. Then, a new unvisited point is retrieved and processed, leading to the discovery of a further cluster or noise.

## 3. Clustering Validity Indices

In this paper the behaviour of several clustering validity indices has been examined (Figure 1). Among these indices some of them are clearly illustrated in this section. These indices are used for measuring "goodness" of a clustering result comparing to other ones which were created by other clustering algorithms or by the same algorithms but using different parameter values [13].

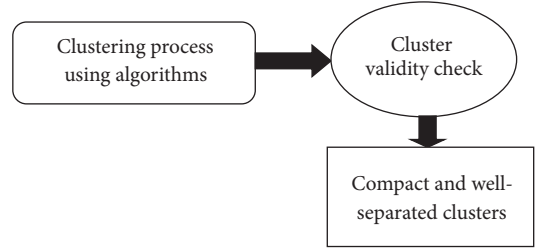

Figure 1: Clustering validity assessment.

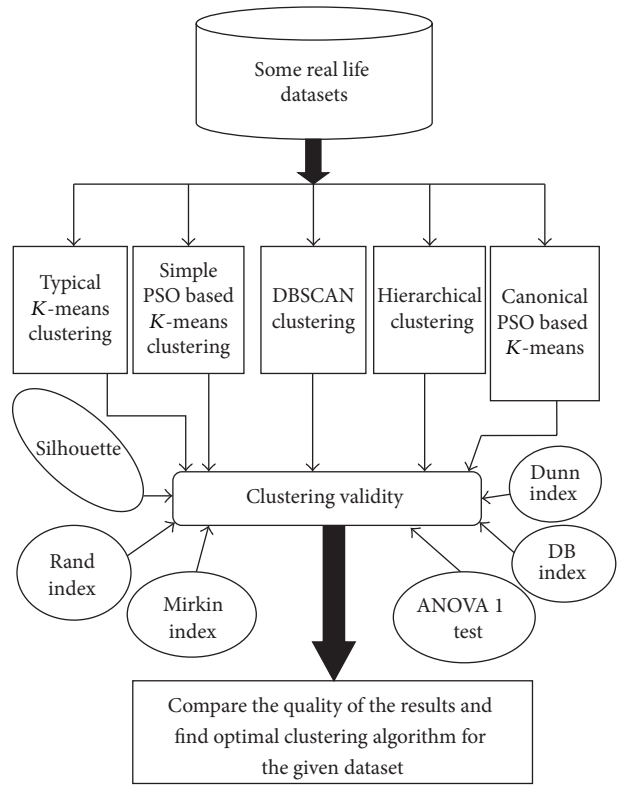

Figure 2: Motivation of this paper.

*3.1. Silhouette Index.* Silhouette index is used for cluster analysis formed by different clustering algorithms (Tables 6, 7, and 8). The silhouette validation technique calculates the silhouette width for each sample, average silhouette width for each cluster, and overall average silhouette width for a total dataset in this paper [14]. The average silhouette width could be applied for evaluation of clustering validity and also could be used to decide how good the number of selected clusters is.

To construct the silhouettes $S(i)$ the following formula is used:

$$s(i) = \frac{(b(i) - a(i))}{\max\{a(i), b(i)\}}, \tag{5}$$

where $a(i)$ is the average dissimilarity of $i$th object to all other objects in the same cluster and $b(i)$ is the minimum of average dissimilarity of $i$-object to all objects in other clusters [9].

If silhouette value is close to 1, it means that sample is "well-clustered" and it was assigned to a very appropriate cluster. If silhouette value is about zero, it means that sample could be assigned to another closest cluster as well, and the sample lies equally far away from both clusters. If silhouette value is close to −1, it means that sample is "misclassified" and is merely somewhere in between the clusters. The overall average silhouette width for the entire plot is simply the average of the $S(i)$ for all objects in the whole dataset.

### 3.2. Davies-Bouldin Index (DB).
The Davies-Bouldin index (DB) can be calculated by the following formula:

$$\text{DB} = \left(\frac{1}{n}\right) \sum_{i=1}^{n} \max_{i \neq j} \left[ \frac{(\partial_i + \partial_j)}{d(C_i, C_j)} \right], \tag{6}$$

where $n$ is the number of clusters, $C_x$ is the centroid of cluster, $\partial_x$ is the average distance of all elements in cluster to centroids, and $d(C_i, C_j)$ is the distance between centroids [15]. Since algorithms that produce clusters with low intracluster distances (high intracluster similarity) and high intercluster distances (low intercluster similarity) will have a low Davies-Bouldin index, the clustering algorithm that produces a collection of clusters with the smallest Davies-Bouldin index is considered the best algorithm based on this criterion [9].

### 3.3. Dunn Index.
The Dunn index aims to identify dense and well-separated clusters. It is defined as the ratio between the minimal intercluster distances to maximal intracluster distance. For each cluster partition, the Dunn index can be calculated by the following formula:

$$D = \min_{1 \leq i \leq n} \left[ \min_{1 \leq j \leq n; i \neq j} \left\{ \frac{d(i,j)}{\max_{1 \leq k \leq n} d'(k)} \right\} \right], \tag{7}$$

where $d(i, j)$ represents the distance between clusters $i$ and $j$ and $d'(k)$ measures the intracluster distance of cluster $k$. The intercluster distance $d(i, j)$ between two clusters may be any number of distance measures, such as the distance between the centroids of the clusters [15]. Similarly, the intracluster distance $d'(k)$ may be measured in a variety of ways, such as the maximal distance between any pair of elements in cluster $k$. Since internal criterion seeks clusters with high intracluster similarity and low intercluster similarity, algorithms that produce clusters with high Dunn index are more desirable [16].

### 3.4. Rand Index.
The Rand index computes how similar the clusters (returned by the clustering algorithm) are to the benchmark classifications. One can also view the Rand index as a measure of the percentage of correct decisions made by the algorithm [16].

Given a set of $n$ elements $S = \{o_1, \ldots, o_n\}$ and two partitions of $S$ to compare, $X = \{X_1, \ldots, X_r\}$, a partition of $S$ into $r$ subsets, and $Y = \{Y_1, \ldots, Y_s\}$, a partition of $S$ into $s$ subsets, then the Rand index, $R$, is

$$R = \frac{a+b}{a+b+c+d} = \frac{a+b}{nC2}, \tag{8}$$

where $a$ is the number of pairs of elements in $S$ that are in the same set in $X$ and in the same set in $Y$, $b$ is the number of pairs of elements in $S$ that are in different sets in $X$ and in different sets in $Y$, $c$ is the number of pairs of elements in $S$ that are in the same set in $X$ and in different sets in $Y$, and $d$ is the number of pairs of elements in $S$ that are in different sets in $X$ and in the same set in $Y$.

### 3.5. Mirkin Index.
Mirkin index is also known as equivalence mismatch distance. It is defined by

$$M(C, C') = \sum_{i=1}^{k} |C_i| + \sum_{j=1}^{l} |C'_j|^2 - 2 \sum_{i=1}^{k} \sum_{j=1}^{k} m_{ij}^2, \tag{9}$$

where $i = 1, \ldots, k$, $j = 1, \ldots, l$.

Let $X$ be a finite set with cardinality $|x| - n$. A clustering $C$ is a set $\{C_1, \ldots, C_k\}$ of nonempty disjoint subsets of $X$ such that their union equals $X$. Clustering $C'$ is a refinement of cluster of $C$, formally:

$$\forall C'_j \in C', \quad \exists C_i \in C : C'_j \subset C_i. \tag{10}$$

### 3.6. ANOVA Test.
"Analysis of variance" test is used to compare three or more groups or conditions in an experiment. A one-way ANOVA can help to find out if the means for each group/condition are significantly different from one another or if they are relatively the same. The null hypothesis typically corresponds to a general or default position. One-way ANOVA is a simple special case of the linear model. The one-way ANOVA form of the model is

$$y_{ij} = \alpha_{.j} + \epsilon_{ij}, \tag{11}$$

where $y_{ij}$ is a matrix of observations in which each column represents a different group. $\alpha_{.j}$ is a matrix whose columns are the group means (the "dot $j$" notation means that applies to all rows of the $j$th column. That is, the value $\alpha_{ij}$ is the same for all $i$). $\epsilon_{ij}$ is a matrix of random disturbances.

The standard ANOVA table has six columns:

(i) the source of the variability,

(ii) the sum of squares (SS) due to each source,

(iii) the degrees of freedom (df) associated with each source,

(iv) the mean squares (MS) for each source, which is the ratio SS/df,

Table 1: Results after first iteration.

| Data objects | Clust1 | Clust2 | Clust3 | New cluster |
|---|---|---|---|---|
| 10 | 0 | 12 | 9 | Clust1 |
| 12 | 2 | 10 | 11 | Clust1 |
| 15 | 5 | 7 | 14 | Clust1 |
| 7 | 3 | 15 | 6 | Clust1 |
| 22 | 12 | 0 | 21 | Clust2 |
| 29 | 19 | 7 | 28 | Clust2 |
| 31 | 21 | 9 | 30 | Clust2 |
| 3 | 7 | 19 | 2 | Clust3 |
| 1 | 9 | 21 | 0 | Clust3 |
| 5 | 5 | 17 | 4 | Clust3 |
| 7 | 3 | 15 | 6 | Clust1 |
| 4 | 6 | 18 | 3 | Clust3 |
| 12 | 2 | 10 | 11 | Clust1 |
| 11 | 1 | 11 | 10 | Clust1 |
| 10 | 0 | 12 | 9 | Clust1 |
| | | | | Number of items |
| Clust1 = {10, 12, 15, 7, 12, 7, 11, 10} | | | | 8 |
| Clust2 = {22, 29, 31} | | | | 3 |
| Clust3 = {3, 1, 5, 4} | | | | 4 |

Table 2: Results after second iteration.

| Data objects | Clust1 | Clust2 | Clust3 | New cluster |
|---|---|---|---|---|
| 10 | 0.5 | 17.3 | 6.75 | Clust1 |
| 12 | 1.5 | 15.3 | 8.75 | Clust1 |
| 15 | 4.5 | 12.3 | 11.75 | Clust1 |
| 7 | 3.5 | 20.3 | 3.75 | Clust1 |
| 22 | 11.5 | 5.3 | 18.75 | Clust2 |
| 29 | 18.5 | 1.7 | 25.75 | Clust2 |
| 31 | 20.5 | 3.7 | 27.75 | Clust2 |
| 3 | 7.5 | 24.3 | 0.25 | Clust3 |
| 1 | 9.5 | 26.3 | 2.25 | Clust3 |
| 5 | 5.5 | 22.3 | 1.75 | Clust3 |
| 7 | 3.5 | 20.3 | 3.75 | Clust1 |
| 4 | 6.5 | 23.3 | 1.25 | Clust3 |
| 12 | 2.5 | 15.3 | 8.75 | Clust1 |
| 11 | 0.5 | 16.3 | 7.75 | Clust1 |
| 10 | 0.5 | 17.3 | 6.75 | Clust1 |
| | | | | Number of items |
| Clust1 = {10, 12, 15, 7, 12, 7, 11, 10} | | | | 8 |
| Clust2 = {22, 29, 31} | | | | 3 |
| Clust3 = {3, 1, 5, 4} | | | | 4 |

(v) the $F$-statistic, which is the ratio of the mean squares,

(vi) the probability value, which is derived from the cdf of $F$.

The ANOVA test makes the following assumptions about the data:

(i) all sample populations are normally distributed;

(ii) all sample populations have equal variance;

(iii) all observations are mutually independent.

## 4. Mathematical Illustration

Every logical experiment needs a well-defined mathematical illustration. In this section, the Davis-Bouldin (DB) index and Dunn index are applied on three most commonly used clustering algorithms (typical $K$-means, DBSCAN, and Hierarchical) with the help of mathematical examples.

Eg.1. Let us assume there is a set of 15 data objects, such as 10, 12, 15, 7, 22, 29, 31, 3, 7, 5, 1, 4, 12, 11, and 10.

Sol:

*K-Means Clustering*

*Step 1.* In the first step typical $K$-means algorithm is applied on this dataset. Suppose the total number of clusters is 3 and also assume their centroids are Clust1 = 10, Clust2 = 22, and Clust3 = 1.

Now use Manhattan distance measure $|d_i - d_j|$ to calculate the following.

First iteration: see Table 1 and Figure 3.
Now, the new centroids are Clust1 = 10.5, Clust2 = 27.3, and Clust3 = 3.25.

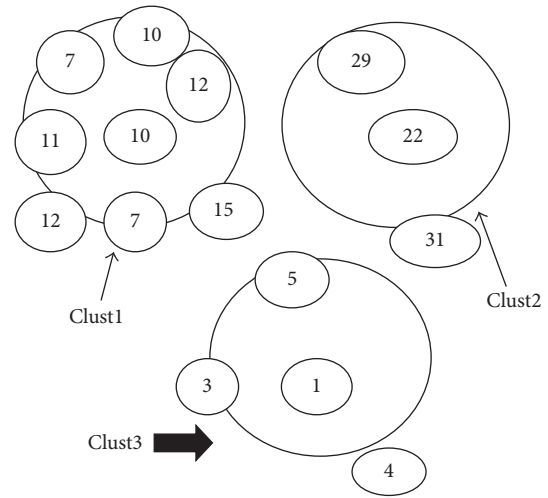

Figure 3: Resultant clusters after the first iteration.

Second iteration: see Table 2.
→No change occurs.
So those three clusters are final clusters and now based on those clusters two popular indexes Davis-Bouldin (DB) index and Dunn index which are calculated below.

*Davies-Bouldin Index.* Consider

$$\text{DB} = \left(\frac{1}{n}\right) \sum_{i=1}^{n} \max_{i \neq j} \left[ \frac{(\partial_i + \partial_j)}{d(C_i, C_j)} \right], \quad (12)$$

Table 3: Dunn index values list.

| Clusters | Dunn index | Values |
|---|---|---|
| Clust1 | $D_{1,2}$ | 0.87 |
|  | $D_{1,3}$ | 1.7 |
| Clust2 | $D_{2,1}$ | 0.77 |
|  | $D_{2,3}$ | 1.88 |
| Clust3 | $D_{3,1}$ | 0.5 |
|  | $D_{3,2}$ | 4.25 |

where $n = 3, C_i = 10, C_j = 22, C_k = 1, \partial_i = \{(0+2+5+3+1+2+3)/8\} = 2, \partial_j = \{(7+9)/3\} = 5.3, \partial_k = \{(2+3+4)/4\} = 2.25, d(C_i, C_j) = 12, d(C_j, C_k) = 21,$ and $d(C_i, C_k) = 9$.

So,

$$\mathrm{DB}(C_i, C_j, C_k)$$
$$= \left(\frac{1}{3}\right) \sum_{i=1}^{3} \max_{i \neq j} [0.61, 0.36, 0.47] \quad (13)$$
$$= \left(\frac{1}{3}\right) \times 0.61$$
$$\mathrm{DB} = 0.203 \ (\text{probably}).$$

*Dunn Index.* The simple form of Dunn index calculation is

$$D = \frac{d_{\min}}{d_{\max}}, \quad (14)$$

where $d_{\min}$ is the minimum distance between two points belonging to different clusters and $d_{\max}$ is the maximum distance between any two points selected from the same cluster [13].

According to this form, the calculated values are as shown in Table 3.

So,

$$\mathrm{Dunn} = \min(0.87, 1.7, 0.77, 1.88, 0.5, 4.25) = 0.5. \quad (15)$$

*DBSCAN Clustering.* Now, on the same dataset DBSCAN clustering algorithm is applied with assuming the value of minimum number of points (minpts) = 3 and the radius of a cluster (eps) = 1 cm. Initially there are three clusters with the same means and they are shown in Figure 4.

According to the DBSCAN clustering criteria, we have the following.

Here two data objects are treated like outliers because both objects 15 and 31 are ≥eps (1 cm) and due to the minimum number points condition does not satisfied by the Clust3 that is why the objects of the clust3 also acts like outliers. But the objects of Clust3 and the objects 15 and 31 all together form a cluster ($C_{\text{new}}$) which satisfies both criteria of DBSCAN clustering. It is also called the cluster of noise.

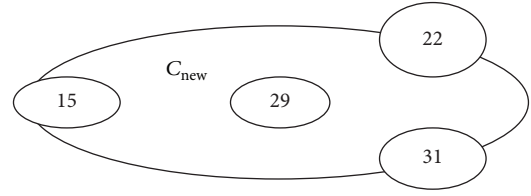

Figure 4: New cluster of outliers.

Now in the same way the DB and DUNN indexes are calculated on those final clusters,

$$\mathrm{DB} = \left(\frac{1}{3}\right) \sum_{i=1}^{3} \max_{i \neq j} (0.39, 0.29, 0.45)$$
$$= 0.25 \ (\text{probably}) \quad (16)$$
$$\mathrm{Dunn} = \min(0.4, 0.6, 1.7, 2.5, 0.63, 0.58)$$
$$= 0.4.$$

These values may be changed based on the assumption of eps and minpts values.

*Evaluation.* After checking the values of the indexes it can be said that for that particular set of data objects the typical $K$-means clustering gives better compactness and better separability than DBSCAN clustering (due to the fact that less value of DB index and high value of Dunn index give more desirable results). But in most of the cases DBSCAN gives bad results compared to $K$-means clustering because it forms arbitrary shape of clusters.

## 5. Result Analysis

*5.1. Datasets*

*5.1.1. Air Pollution Dataset.* The experimented air pollution database is taken from the http://www.wbpcb.gov.in/ website of Kolkata District from 01/07/2011 to 01/05/2012. The database contains different air pollutant data like benzene, $CO_2$ (carbon-dioxide), $NO_2$ (nitrogen-dioxide), $O_3$ (ozone), PM (particular matter), and $SO_2$ (sulphur-dioxide).

*5.1.2. Customer Dataset.* The wholesale customer dataset is taken from UCI Machine Learning Repository. The dataset contains eight attributes like fresh products, milk, grocery, frozen, detergents, delicatessen, channel, and region.

*5.1.3. Wine Dataset.* The wholesale wine dataset is taken from UCI Machine Learning Repository. The dataset contains thirteen attributes like alcohol, malic acid, ash, alkalinity of ash, magnesium, total phenols, flavanoids, nonflavanoid phenols, proanthocyanins, color intensity, hue, OD280/OD315 of diluted wines, and proline.

*5.1.4. Vehicle Dataset.* The wine dataset is taken from UCI Machine Learning Repository. The dataset contains eighteen attributes like compactness, circularity, distance circularity area, radius ratio, Pr.Axis aspect ratio, max. length aspect ratio, scatter ratio, elongatedness, Pr.Axis rectangularity, max. length rectangularity area, scaled variance along major axis, scaled variance along minor axis, scaled radius of gyration, skewness about major axis, skewness about minor axis, kurtosis about minor axis, kurtosis about major axis, and hollows ratio. Here 4 classes are considered: Opel, Saab, Bus, and Van. The 4 real datasets and their main characteristics are shown in "Table 4." The experiment is based on 150 iterations and five clustering algorithms.

Table 4: The characteristics of real datasets drawn from UCI repository.

| Dataset | Number of objects | Features | Classes |
| --- | --- | --- | --- |
| Air pollution | 305 | 7 | 5 |
| Customer | 440 | 8 | 2 |
| Wine | 178 | 13 | 3 |
| Vehicle | 846 | 18 | 3 |

Table 5: Silhouette index comparison of typical $K$-means, DBSCAN Hierarchical, simple PSO based $K$-means, and Canonical PSO based $K$-means algorithm on air pollution dataset.

| Typical $K$-means | DBSCAN clustering | Hierarchical clustering | Simple PSO based $K$-means | Canonical PSO based $K$-means |
| --- | --- | --- | --- | --- |
| 0.5445 | 0.5379 | 0.6345 | 0.5813 | 0.6556 |

Table 6: Silhouette index comparison of typical $K$-means, DBSCAN, Hierarchical, simple PSO based $K$-means, and Canonical PSO based $K$-means algorithm on customer dataset.

| Typical $K$-means | DBSCAN clustering | Hierarchical clustering | Simple PSO based $K$-means | Canonical PSO based $K$-means |
| --- | --- | --- | --- | --- |
| 0.6330 | 0.7335 | 0.3272 | 0.6869 | 0.8006 |

Table 7: Silhouette index comparison of typical $K$-means, DBSCAN, Hierarchical, simple PSO based $K$-means, and Canonical PSO based $K$-means algorithm on wine dataset.

| Typical $K$-means | DBSCAN clustering | Hierarchical clustering | Simple PSO based $K$-means | Canonical PSO based $K$-means |
| --- | --- | --- | --- | --- |
| 0.7323 | 0.7150 | 0.7022 | 0.8060 | 0.9003 |

Table 8: Silhouette index comparison of typical $K$-means, DBSCAN, Hierarchical, simple PSO based $K$-means, and Canonical PSO based $K$-means algorithm on vehicle dataset.

| Typical $K$-means | DBSCAN clustering | Hierarchical clustering | Simple PSO based $K$-means | Canonical PSO based $K$-means |
| --- | --- | --- | --- | --- |
| 0.7240 | 0.1354 | 0.4600 | 0.7217 | 0.8469 |

5.2. Experimental Analysis. At first analysis, the accuracy of different clustering method is measured using silhouette index depicted in Tables 2, 3, 4, and 5. Probably 150 iterations are considered in each case. One-way ANOVA test is applied on the values of silhouette index at different iterations and null hypothesis has been measured and reported in Tables 9, 10, 11, and 12. The accuracy comparisons of all clustering algorithms on different datasets are explained in Table 13.

Table 9: ANOVA 1 test for $K$-means, DBSCAN, Hierarchical, simple PSO based $K$-means, and Canonical PSO based $K$-means algorithm on air pollution dataset.

| | ANOVA table | | | | |
| --- | --- | --- | --- | --- | --- |
| Source | SS | df | MS | F | Prob > F |
| Columns | 0.0323 | 4 | 0.00807 | | |
| Error | 0.00023 | 5 | 0.00005 | 176.91 | $1.43598e - 005$ |
| Total | **0.03253** | **9** | | | |

The results of ANOVA test is as shown in Tables 9, 10, 11, and 12 and Figures 5, 6, 7, and 8.

The reason for doing an ANOVA test is to see if there is any difference between groups on some variables. If probability $> F$, we reject null hypothesis. Here in all cases, Prob $> F$. Therefore, we reject the null hypothesis. The means of the five algorithms are not equal. At least one of the means is different. However, the ANOVA test does not tell where the difference lies. It requires a "$t$-test" to test each pair of means.

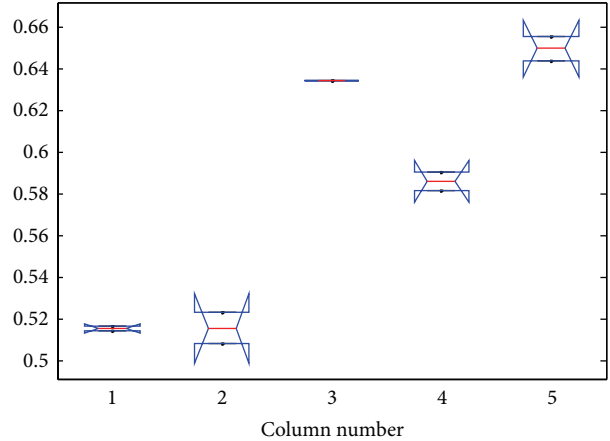

Figure 5: The box plot view of the ANOVA 1 test for air pollution dataset.

Table 10: ANOVA 1 test for $K$-means, DBSCAN, Hierarchical, simple PSO based $K$-means, and Canonical PSO based $K$-means algorithm for customer dataset.

| | ANOVA table | | | | |
|---|---|---|---|---|---|
| Source | SS | df | MS | F | Prob > F |
| Columns | 0.38985 | 4 | 0.09746 | | |
| Error | 0.00772 | 10 | 0.00077 | 126.17 | $1.6346e - 008$ |
| Total | **0.39758** | **14** | | | |

Table 11: ANOVA 1 test for $K$-means, DBSCAN, Hierarchical, simple PSO based $K$-means, and Canonical PSO based $K$-means algorithm for wine dataset.

| | ANOVA table | | | | |
|---|---|---|---|---|---|
| Source | SS | df | MS | F | Prob > F |
| Columns | 0.05843 | 4 | 0.01461 | | |
| Error | 0.00343 | 10 | 0.00034 | 42.59 | $2.99893e - 006$ |
| Total | **0.06186** | **14** | | | |

Table 12: ANOVA 1 test for $K$-means, DBSCAN, Hierarchical, simple PSO based $K$-means, and Canonical PSO based $K$-means algorithm for vehicle dataset.

| | ANOVA table | | | | |
|---|---|---|---|---|---|
| Source | SS | df | MS | F | Prob > F |
| Columns | 0.80774 | 4 | 0.20193 | | |
| Error | 0.02507 | 10 | 0.00251 | 80.55 | $1.44581e - 007$ |
| Total | **0.83281** | **14** | | | |

## 6. Discussion

So from Table 13 it can be concluded that Canonical PSO based $K$-means algorithm provides most desirable results compared to other clustering algorithms stated in Section 2 and DBSCAN clustering provides the worst results than the others (based on the initial values of eps = 25 and minpts = 65). In this part, the results of these algorithms are evaluated by applying other well-known indexes (such as, Dunn, Davies-Bouldin, Rand, and Mirkin indices).

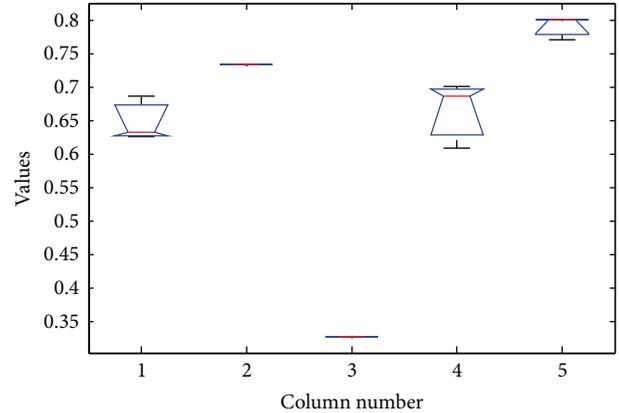

Figure 6: The box plot view of the ANOVA 1 test for customer dataset.

From "Table 14" the results of this experiment can be summerised as follows.

(1) According to the Dunn index, algorithm that produces clusters with high Dunn index is more desirable. So here the Canonical PSO based $K$-means algorithm produces better results compared to the others (0.7132 > 0.6542 > 0.6152 > 0.5922 > 0.3984).

(2) According to the Davis-Bouldin (DB) index, algorithm that produces a collection of clusters with the smallest Davies-Bouldin index is considered the best algorithm. So here again the Canonical PSO based $K$-means algorithm produces better results compared to the others (0.2074 < 0.3283 < 0.3344 < 0.5484 < 0.5816).

(3) According to the Rand index, algorithm that produces clusters with high Rand index is more suitable and compact. So here the Canonical PSO based $K$-means algorithm produces more good results than the others (0.7825 > 0.6812 > 0.4687 > 0.3702 > 0.3022).

Table 13: Accuracy comparison of all datasets.

| Dataset | Typical K-means | DBSCAN clustering | Hierarchical clustering | Simple PSO based K-means | Canonical PSO based K-means |
|---|---|---|---|---|---|
| Air pollution | 0.5652 | 0.4880 | 0.5452 | 0.6882 | 0.7872 |
| Customer | 0.5174 | 0.5678 | 0.5581 | 0.5845 | 0.6880 |
| Wine | 0.7202 | 0.4147 | 0.7161 | 0.8061 | 0.8294 |
| Vehicle | 0.5417 | 0.2599 | 0.5633 | 0.6965 | 0.7217 |

Table 14: Various indices comparison based on the air pollution dataset.

| Indices | Typical K-means | DBSCAN clustering | Hierarchical clustering | Simple PSO based K-means | Canonical PSO based K-means |
|---|---|---|---|---|---|
| Dunn | 0.6542 | 0.3984 | 0.5922 | 0.6152 | 0.7132 |
| DB | 0.5484 | 0.5816 | 0.3283 | 0.3344 | 0.2074 |
| Rand | 0.3022 | 0.3702 | 0.4687 | 0.6812 | 0.7825 |
| Mirkin | 0.7167 | 0.3451 | 0.4802 | 0.2312 | 0.1123 |

Table 15: Various indices comparison based on customer dataset.

| Indices | Typical K-means | DBSCAN clustering | Hierarchical clustering | Simple PSO based K-means | Canonical PSO based K-means |
|---|---|---|---|---|---|
| Dunn | 0.3451 | 0.5327 | 0.4543 | 0.5734 | 0.6258 |
| DB | 0.331 | 0.2873 | 0.3952 | 0.2342 | 0.1453 |
| Rand | 0.4642 | 0.5045 | 0.4889 | 0.6112 | 0.6953 |
| Mirkin | 0.5228 | 0.4380 | 0.4375 | 0.3421 | 0.3166 |

Table 16: Various indices comparison based on the wine dataset.

| Indices | Typical K-means | DBSCAN clustering | Hierarchical clustering | Simple PSO based K-means | Canonical PSO based K-means |
|---|---|---|---|---|---|
| Dunn | 0.4003 | 0.5368 | 0.5443 | 0.5368 | 0.6953 |
| DB | 0.4962 | 0.3491 | 0.1879 | 0.2432 | 0.2312 |
| Rand | 0.3759 | 0.4699 | 0.2828 | 0.4598 | 0.4624 |
| Mirkin | 0.6240 | 0.5376 | 0.5921 | 0.3032 | 0.2948 |

Table 17: Various indices comparison based on the vehicle dataset.

| Indices | Typical K-means | DBSCAN clustering | Hierarchical clustering | Simple PSO based K-means | Canonical PSO based K-means |
|---|---|---|---|---|---|
| Dunn | 0.5652 | 0.5654 | 0.4332 | 0.4899 | 0.5256 |
| DB | 0.2329 | 0.1987 | 0.2591 | 0.1899 | 0.1754 |
| Rand | 0.4321 | 0.7296 | 0.5377 | 0.6123 | 0.6417 |
| Mirkin | 0.2159 | 0.2259 | 0.3140 | 0.2910 | 0.1512 |

(4) Mirkin index gives smallest value when the number of clusters attains optimal level. So here Canonical PSO based K-means algorithm produces much better results compared to the others (0.1123 < 0.2312 < 0.3451 < 0.4802 < 0.7167). But the inefficient clustering algorithm in this experiment is not fixed according to the above indices of Table 14. But after the average cases analysis, it can be noticed that the DBSCAN algorithm produces improper clusters than the others (but it totally based on the initial assumption of minpts and eps values of DBSCAN algorithm). All these results and analysis are represented in a pictorial form as in Figure 9.

The same approach and results are also valid for the other 3 datasets (such as customer, wine, and vehicle). The detailed analysis of results is tabulated in Tables 15, 16, and 17 and their corresponding pictorial forms are represented in Figures 10, 11, and 12.

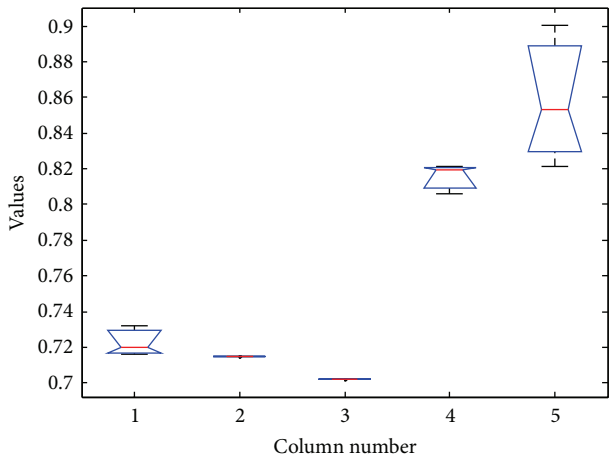

Figure 7: The box plot view of the ANOVA 1 test for wine dataset.

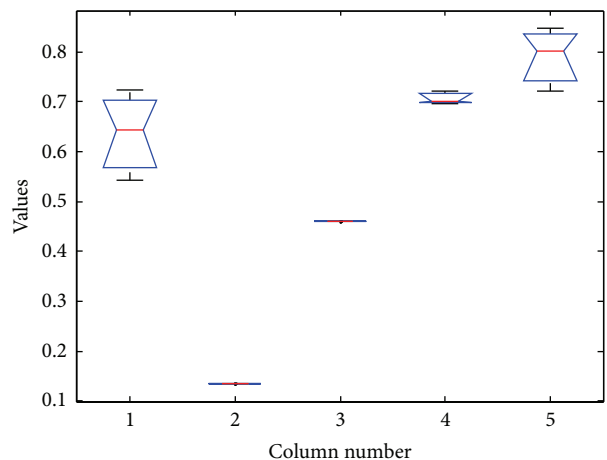

Figure 8: The box plot view of the ANOVA 1 test for vehicle dataset.

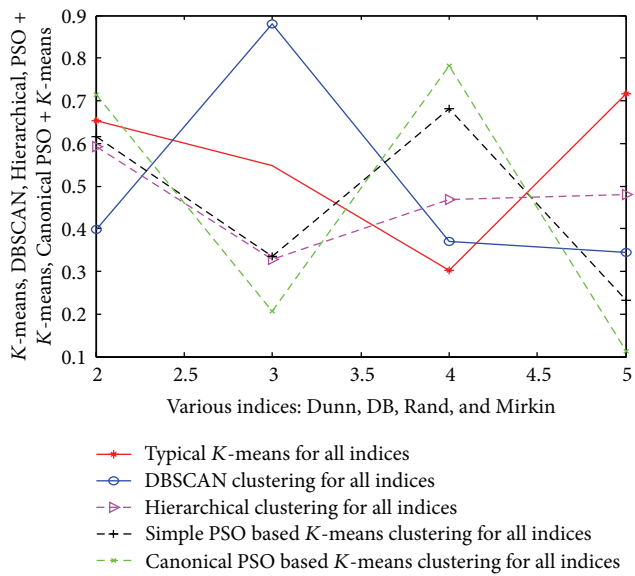

Figure 9: Representation of various indices using air pollution dataset.

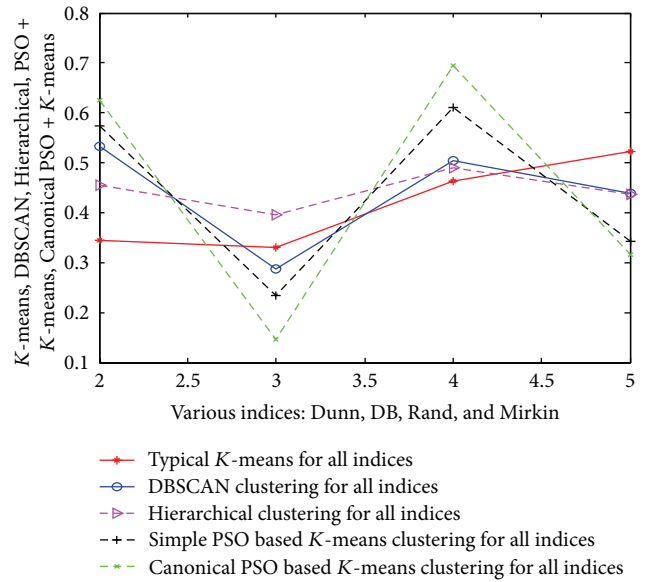

Figure 10: Representation of various indices using customer dataset.

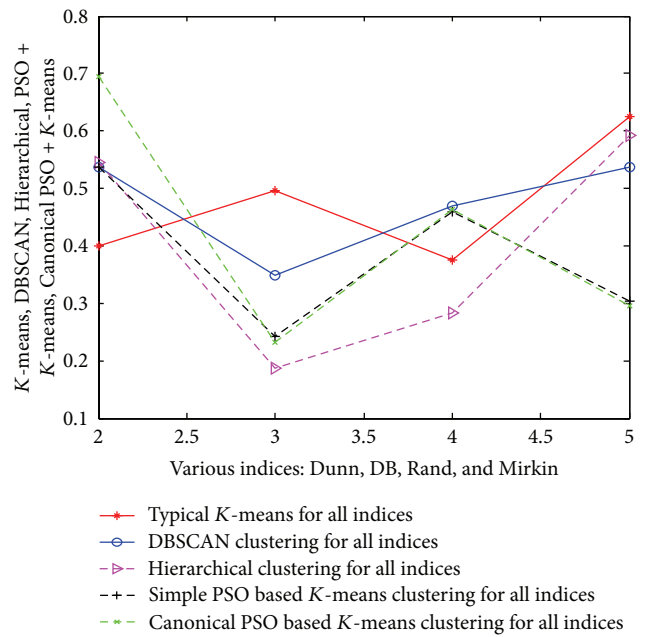

Figure 11: Representation of various indices using wine dataset.

## 7. Conclusion and Future Work

Although there are several proposals for validity indices in the literature, most of them succeed only in certain situations and are based on some certain conditions. This paper introduces the new concept like Canonical PSO based $K$-means algorithm and also presents ample comparisons of six popular clustering validity approaches. All these analyses are done on a few real time datasets which contain air pollutant particles, wholesale customer, wine, and vehicle data. This proposed algorithm provides better impact to produce desirable clustering results compared to other existing algorithms described in this paper (provided some certain assumptions).

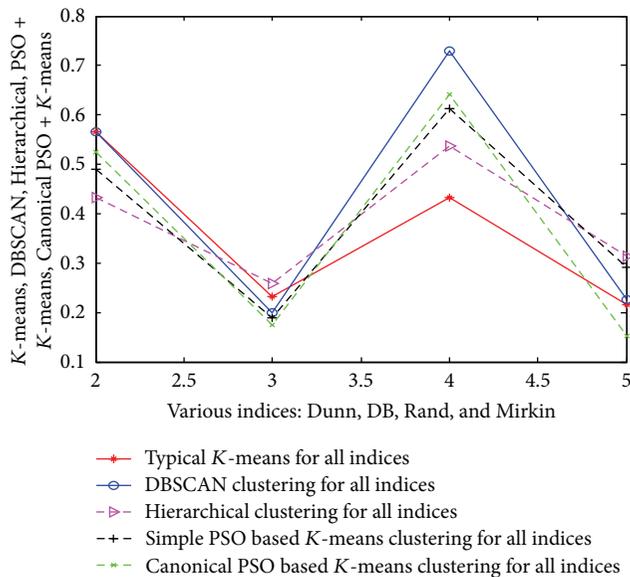

Figure 12: Representation of various indices using vehicle dataset.

Several extensions to the present work are in progress. Finally, studies of other clustering algorithms and others indices and their effects to measure proper, compact clusters are also underway. In future work, we can amalgamate DBSCAN and Hierarchical algorithms with PSO approach to seek some better ways of producing improved clustering output.

## Conflict of Interests

The authors declare that there is no conflict of interests regarding the publication of this paper.

## References

[1] S. Saitta, B. Raphael, and I. F. C. Smith, "A bounded index for Cluster validity," in *Proceedings of the 5th International Conference on Machine Learning and Data Mining in Pattern Recognition*, pp. 174–187, Springer, 2007.

[2] L. Jegatha Deborah, R. Baskaran, and A. Kannan, "A survey on internal validity measure for cluster validation," *International Journal of Computer Science & Engineering Survey*, vol. 1, no. 2, pp. 85–102, 2010.

[3] E. Rendón, I. M. Abundez, and C. Gutierrez, "A comparison of internal and external cluster validation indexes," in *Proceedings of the 5th WSEAS International Conference on Computer Engineering and Applications*, pp. 158–163, 2011.

[4] F. Kovács, C. Legány, and A. Babos, "Cluster validity measurement techniques," in *Proceedings of the 5th International Conference on Artificial, Intelligence, Knowledge Engineering and Data Bases (WSEAS '06)*, pp. 388–393, 2006.

[5] G. Stegmayer, D. H. Milone, L. Kamenetzky, M. G. Lopez, and F. Carrari, "A biologically inspired validity measure for comparison of clustering methods over metabolic data sets," *IEEE/ACM Transactions on Computational Biology and Bioinformatics*, vol. 9, no. 3, pp. 706–716, 2012.

[6] K. Premalatha and A. M. Natarajan, "A new approach for data clustering based on PSO with local search," *Computer and Information Science*, vol. 1, no. 4, pp. 139–145, 2008.

[7] X. Cui and T. E. Potok, "Document clustering analysis based on hybrid PSO+K-means algorithm," *Journal of Computer Sciences*, pp. 27–33, 2005.

[8] X. Cui, T. E. Potok, and P. Palathingal, "Document clustering using particle swarm optimization," in *Proceedings of the IEEE Swarm Intelligence Symposium (SIS '05)*, pp. 185–191, Pasadena, Calif, USA, June 2005.

[9] L. Dey and A. Mukhopadhyay, "Microarray gene expression data clustering using PSO based K-means algorithm," in *Proceedings of the International Conference on Advanced Computing, Communication and Networks*, vol. 1, pp. 587–591, 2011.

[10] K. Dhanalakshmi and H. Hannah Inbarani, "Fuzzy soft rough K-Means clustering approach for gene expression data," *International Journal of Scientific and Engineering Research*, vol. 3, no. 10, 2012.

[11] DBSCAN, http://en.wikipedia.org/wiki/DBSCAN.

[12] M. Ester, H.-P. Kriegel, J. Sander, and X. Xu, "A density-based algorithm for discovering clusters in large spatial databases with noise," in *Proceedings of the 2nd International Conference on Knowledge Discovery and Data Mining (KDD '96)*, 1998.

[13] N. Speer, C. Spieth, and A. Zell, "Biological cluster validity indices based on the gene ontology," in *Advances in Intelligent Data Analysis VI*, vol. 3646 of *Lecture Notes in Computer Science*, pp. 429–439, Springer, Berlin, Germany, 2005.

[14] D. Jiang, C. Tang, and A. Zhang, "Cluster analysis for gene expression data: a survey," *IEEE Transactions on Knowledge and Data Engineering*, vol. 16, no. 11, pp. 1370–1386, 2004.

[15] "Cluster Analysis," http://en.wikipedia.org/wiki/Cluster_analysis.

[16] Z. Ansari, M. F. Azeem, W. Ahmed, and A. Vinaya Babu, "Quantitative evaluation of performance and validity indices for clustering the web navigational sessions," *World of Computer Science and Information Technology Journal*, vol. 1, no. 5, pp. 217–226, 2011.

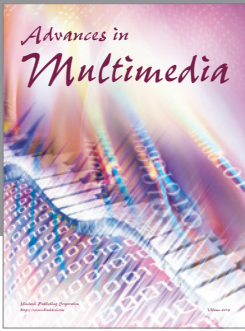
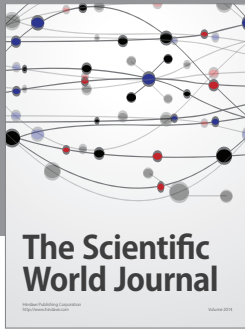
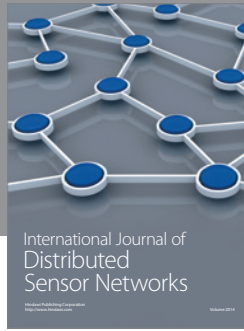
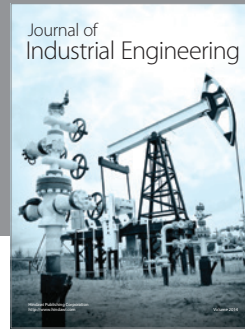
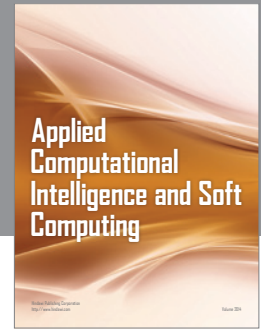
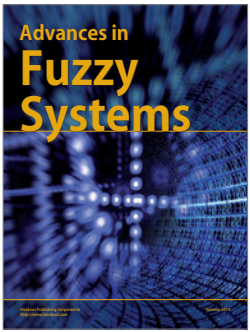
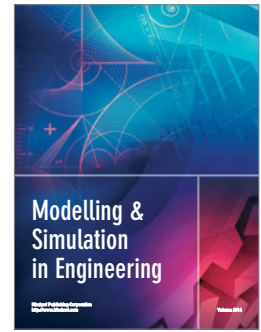
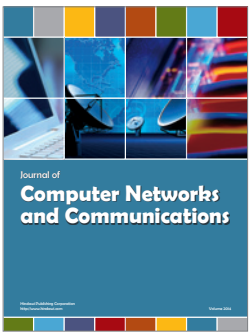
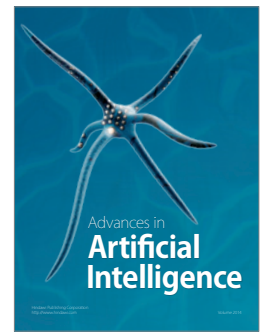
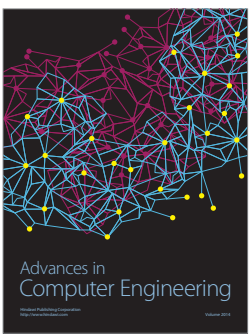
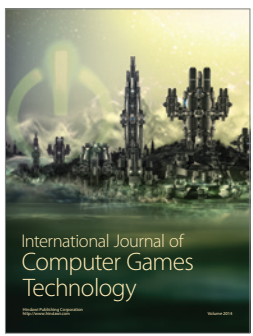
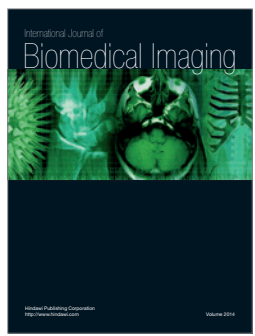
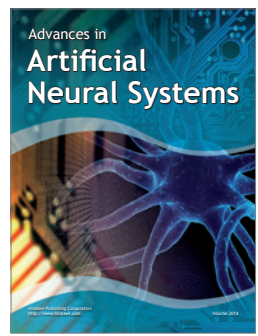
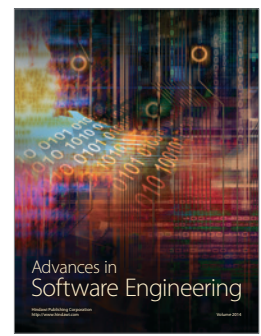
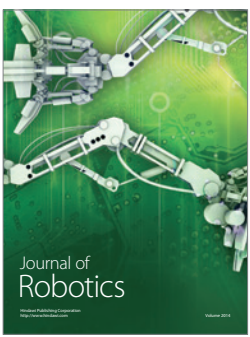
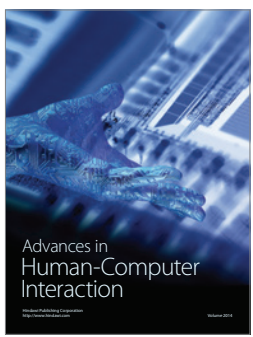
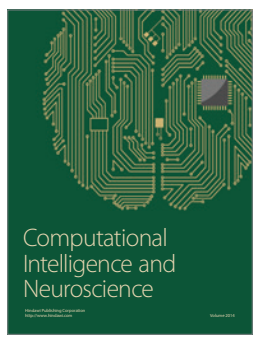
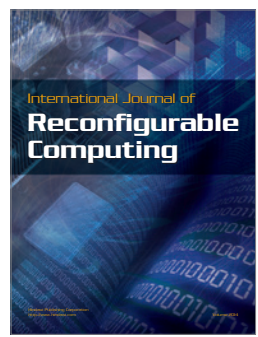
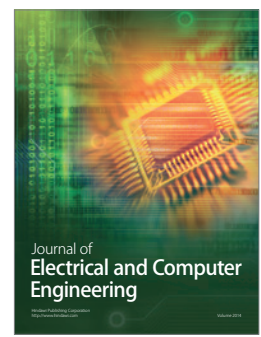